\documentclass[twocolumn,trackchanges]{aastex63}

\newcommand{\hst}{{\it HST}}

\newcommand{\mbh}{$\mathcal M_{\rm BH}$}

\newcommand{\mr}{$Mag_{\rm ~R}$}
\newcommand{\halpha}{${\it H}\alpha$}

\newcommand{\sersic}{S\'ersic}
\newcommand{\lenstronomy}{{\sc Lenstronomy}}
\newcommand{\reff}{{$R_{\mathrm{eff}}$}}
\newcommand{\kmsMpc}{km~s$^{\rm -1}$~Mpc$^{\rm -1}$}
\newcommand{\kms}{\ifmmode{\,\rm{km}\, \rm{s}^{-1}}\else{$\,$km$\,$s$^{-1}$}\fi}

\newcommand{\mstar}{{$M_*$}}
\newcommand{\Mgii}{Mg$_{\rm II}$}
\newcommand{\Civ}{C$_{\rm IV}$}
\newcommand{\sam}{\texttt{SAM}}
\newcommand{\mbii}{\texttt{MBII}}

\shorttitle{Confronting co-evolution theory with observations}
\shortauthors{Ding et al.}

\begin{document}

\title{Testing the fidelity of simulations of black hole - galaxy co-evolution at $z\sim1.5$ with observations}

\correspondingauthor{Xuheng Ding}
\email{dxh@astro.ucla.edu}

\author[0000-0001-8917-2148]{Xuheng Ding}
\affiliation{Department of Physics and Astronomy, University of California, Los Angeles, CA, 90095-1547, USA} 

\author[0000-0002-8460-0390]{Tommaso Treu}
\affiliation{Department of Physics and Astronomy, University of California, Los Angeles, CA, 90095-1547, USA} 

\author[0000-0002-0000-6977]{John D. Silverman}
\affiliation{Kavli Institute for the Physics and Mathematics of the Universe (WPI), The University of Tokyo, Kashiwa, Chiba 277-8583, Japan} 
\affiliation{Department of Astronomy, School of Science, The University of Tokyo, 7-3-1 Hongo, Bunkyo, Tokyo 113-0033, Japan}

\author[0000-0002-7080-2864]{Aklant K. Bhowmick}
\affiliation{McWilliams Center for Cosmology, Dept. of Physics, Carnegie Mellon University, Pittsburgh PA 15213, USA}
\affiliation{Department of Physics, University of Florida, Gainesville, FL 32611, USA}

\author[0000-0002-4096-2680]{N. Menci}
\affiliation{INAF Osservatorio Astronomico di Roma, via Frascati 33, I-00078 Monteporzio, Italy}

\author[0000-0002-6462-5734]{Tiziana Di Matteo}
\affiliation{McWilliams Center for Cosmology, Dept. of Physics, Carnegie Mellon University, Pittsburgh PA 15213, USA}

\begin{abstract}
We examine the scaling relations between the mass of a supermassive black hole (SMBH) and its host galaxy properties at $1.2<z<1.7$ using both observational data and simulations. Recent measurements of 32 X-ray-selected broad-line Active Galactic Nucleus (AGNs) are compared with two independent state-of-the-art efforts, including the hydrodynamic simulation \texttt{MassiveBlackII} (\mbii) and a semi-analytic model (\sam).  After applying an observational selection function to the simulations, we find that both \mbii\ and \sam\ agree well with the data, in terms of the central distribution. However, the dispersion in the mass ratio between black hole mass and stellar mass is significantly more consistent with the \mbii\ prediction ($\sim0.3$~dex), than with the \sam\ ($\sim0.7$~dex), even when accounting for observational uncertainties. Hence, our observations can distinguish between the different recipes adopted in the models. The mass relations in the \mbii\ are highly dependent on AGN feedback while the relations in the \sam\ are more sensitive to galaxy merger events triggering nuclear activity. Moreover, the intrinsic scatter in the mass ratio of our high-$z$ sample is comparable to that observed in the local sample, all but ruling out the proposed scenario the correlations are purely stochastic in nature arising from some sort of cosmic central limit theorem. Our results support the hypothesis of AGN feedback being responsible for a causal link between the SMBH and its host galaxy, resulting in a tight correlation between their respective masses.

\end{abstract}



\section{Introduction} \label{sec:intro}
Supermassive black holes (SMBHs) ubiquitously occupy the center of massive galaxies in the local Universe and beyond. Their growth in mass (\mbh) appears to be closely linked to the physical properties of their host galaxies, in particular the relation between \mbh ~and stellar mass (\mstar)~\citep{Mag++98, F+M00, M+H03, H+R04, Gul++09}, indicating a physical coupling during their co-evolution.
Various models have been proposed to explain this connection between SMBH and their host galaxies. A possible physical link may be feedback from an Active Galactic Nucleus (AGN) phase, assuming that a fraction of the AGN energy is injected into their surrounding gas thus regulating the growth of the SMBH and its host galaxy. In this scenario, AGN activity heats and unbinds a significant fraction of the gas and inhibits star formation. An alternative and more indirect connection is one where AGN accretion and star formation are fed through a common gas supply~\citep{Cen2015, Menci2016}. A completely different view holds that the statistical convergence from galaxy assembly alone (i.e., dry mergers) may reproduce the observed correlations without any direct physical mechanisms~\citep{Peng2007, Jahnke2011, Hirschmann2010}. From this central limit theorem, a stochastic cloud at high-$z$ (higher dispersion) would end up with scaling relations as observed today with lower dispersion.

Considerable efforts have been undertaken to establish the scaling relations out to high redshift ($z \lesssim2$) using the {\it Hubble Space Telescope} (\hst) to detect the host galaxies of AGN \citep[e.g.,][]{Peng2006a, Tre++07, Woo++08,Jahnke2009,Bennert11,Schramm2013,Park15,Mechtley2016,Ding2020}. While some studies find an {\it observed} evolution in which the growth of SMBHs predates their host galaxies, there are equally as many claims of no evolution when considering the total stellar mass of the host. For all studies, an understanding of the systematic uncertainties and selection effects~\citep{Tre++07,Lauer2007,Schulze2014} need to be considered to avoid an {\it apparent} evolution that may overestimate the significance of the evolution~\citep{Volonteri2011}.


Simulations can effectively aid in our understanding of this connection by ruling out theories and assumptions that could not be definitively verified by observations alone. In particular, simulations can be used to quantify the impact of systematic uncertainties and selection biases with observational data. For example, the state-of-the-art cosmological hydrodynamical simulation of structure formation (\texttt{MassiveBlackII}) has been used to compare the predicted scaling relations to \hst\ observation at $0.3<z<1$ which show a positive evolution where the SMBH growth predates that of its the host galaxy~\citep{DeG++15}. Several other works have investigated scaling relations using large-volume simulations, resulting in good agreement with the local relation and some redshift evolution, including the Magneticum Pathfinder SPH Simulations~\citep{Steinborn2015}, the Evolution and Assembly of GaLaxies and their Environments (EAGLE) suite of SPH simulations~\citep{Schaye2015}, Illustris moving mesh simulation~\citep{Sijacki2015, Vogelsberger2014, Li2019} and SIMBA simulation~\citep{Thomas2019}. Besides hydrodynamic simulations, semi-analytic models~\citep[e.g.,][]{Menci2014, Menci2016} have also made remarkable progress and recovered the local scaling relations~\citep{Kormendy13}. These comparisons between simulations and the observed scaling relations, based on high-resolution \hst\ imaging, have been carried out mainly at $z<1$ due to prior limitations of the availability of observational data.

In this study, we directly compare the mass ratios between SMBHs and their host galaxies of 32 type-1 (broad-line) AGNs from our recent observational study at $z\sim1.5$ \citep{Ding2020} to the predictions based on two independent state-of-the-art numerical simulations. The redshift range of our targets is chosen ($z<2$) to coincide with the epoch when most of the SMBHs acquired their mass. This choice minimizes sensitivity to initial conditions and growth mechanisms, which in turn allows for an identification of inaccurate or missing physics in the models. This redshift range is also low enough to limit the effect of surface brightness dimming that would lower the success rate of detecting the underlying host galaxy with \hst. We describe the observed and simulated galaxies and their black holes in Section~\ref{sec:sample}. The comparisons between data and simulations are shown in Section~\ref{sec:result} and conclusion presented in Section~\ref{sec:conclusion}. Throughout this paper, we adopt a standard concordance
cosmology with $H_0 = 70$~\kmsMpc
, $\Omega_m = 0.30$,
and $\Omega_\Lambda = 0.70$. A Salpeter initial mass function is employed consistently to the
observed and simulated sample.

\section{Sample: Observations and Simulations}~\label{sec:sample}
In this section, we introduce our comparison samples, including the observed scaling relations (Section~\ref{hst_sample}) and the predicted ones by two independent numerical simulations (Section~\ref{sample_sim}).
	
\subsection{\hst\ Observational data}\label{hst_sample}
	
We have been constructing and analyzing a sample of 32 \hst-observed AGN systems over the redshift range $1.2<z<1.7$ from three deep survey fields, namely COSMOS~\citep{Civano2016}, (E)-CDFS-S~\citep{Lehmer2005, Xue2011}, and SXDS~\citep{Ueda2008} \citep[][hereafter D20]{Ding2020}.
We selected our AGN sample in a well-defined window based on the \mbh\ and Eddington ratio, as shown in Figure~1 in D20. As shown in that Figure, the \mbh\ are well below the knee of the BH mass distribution to avoid a strong selection bias. In addition, the Eddington ratios are mostly above $0.1$ to ensure homogeneity.

We measure reliable \mbh\ and host properties (i.e., \mstar) with a quantitative assessment of systematic effects. Specifically, \mbh\ is determined using published near-infrared spectroscopic observations of the broad \halpha\ emission line, which eliminates potential systematic uncertainties that may arise from switching between Balmer lines in the local universe to the \Mgii\ or \Civ\ UV lines for distant galaxies. Regarding the detection of the host galaxies, the X-ray selected nature of the sample results in slightly lower nuclear-to-host ratios, which facilitates the inference of the host light. 

To detect their host galaxies, we used \hst/WFC3 to obtain high-resolution (0\farcs0642 per pixel) infrared imaging data for 32 AGN systems (HST Program GO-15115). The filters F125W $(1.2<z<1.44)$ and F140W $(1.44<z<1.7)$ were employed, according to the redshift of each target. Six dithered exposures with a total exposure time $\sim2394s$ were co-added using the {\sc astrodrizzle} software package to generate a final image with a pixel scale as 0\farcs0642. We implemented the {\sc photuils} tools to remove contamination due to background light from both the sky and the detector. A full detailed description of the \hst\ data analysis can be found in a companion paper (D20).

We implemented state-of-the-art techniques to perform 2D flux profile decomposition to disentangle the host emission from the AGN. To address biases with respect to the accuracy of the point-spread function (PSF), we collected 2D profiles of isolated and unsaturated stars from the 32 observed \hst\ fields to assemble a PSF library for the fitting routine. To decompose each AGN image, we assume the unresolved active nuclei as scaled point source and the host galaxy as a \sersic\ profile. We ran the imaging modeling tool \lenstronomy~\citep{lenstronomy} to simultaneously fit their 2D flux distribution, taking each PSF one-by-one from the collected PSF library. Based on the reduced $\chi^2$, we are capable of evaluating the performance of each PSF. We adopt the result from the top-ranked-eight PSFs and used a weighting process to obtain the host properties, including flux, \reff, \sersic\ index.

The COSMOS survey provides \hst\ ACS/F814W imaging data for 21 of 32 AGN in our sample with a drizzled pixel scale as 0\farcs03. We decompose the AGN image in the ACS band to obtain the host flux and compare to the WFC3 result to infer the host color. We find that the $1$~Gyr and $0.625$~Gyr stellar templates could well match the sample color at $z<1.44$ and $z>1.44$ respectively, see Figure~5 in D20, from which we estimate the colors of the host to derive the rest frame R-band magnitude (\mr) and the stellar mass (\mstar)-to-light ratio.

We remark that the observational data used in this study is limited to our carefully-constructed sample with a well-understood selection function and 2D assessment of the host galaxy emission from space-based imaging. While there exists larger data sets such as \citet{Sun2015} and others, we refrain from including cases where the host galaxy was assessed using fitting of the spectral energy distribution~(SED) with broad-band photometry from the ground since there are likely higher levels of uncertainty with respect to cases with an AGN of considerable luminosity. However, we recognize that there has yet been definitive evidence for inherent problems with these methods.

\subsection{Numerical simulations}\label{sample_sim}
To compare with simulations, we use two independent efforts, the \texttt{MassiveBlackII} (\texttt{MBII})~\citep{Khandai2015} and the semi-analytic model (\texttt{SAM})~\citep{Menci2014}. These simulations are based on independent model strategies, i.e., hydrodynamic simulation for \mbii\ and semi-analytic model for \sam, respectively.

The \mbii\ simulation is the highest resolution at the size of a comoving volume $V_{\rm box} = (100~{\rm Mpc}~h^{-1})$, including a self-consistent model for star formation, black hole accretion, and associated feedback. The large simulation volume enables the Fourier density modes on the largest scales to evolve independently; the large dynamic range in mass and high spatial resolution meet the requirements to study individual galaxies. While high-resolution N-body simulations can describe specific galaxy systems, an understanding of the physical mechanisms influencing the scaling relations require an analytical description of such processes to be implemented into existing semi-analytic models including the \sam. In previous works, \mbii~\citep{Huang2018, DeG++15, Khandai2015,Bhowmick2019} and \sam~\citep{Menci2014, Menci2016} have made highly successful predictions. In the following two sections, we present detailed information on the two simulation projects.

\subsubsection{\texttt{MassiveBlackII} simulation}
MassiveBlackII (\mbii) is a high-resolution cosmological hydrodynamic simulation using Smooth Particle Hydrodynamics~(SPH) code \texttt{P-GADGET}, which is an upgraded version of~\texttt{GADGET-2}~\citep{2005MNRAS.364.1105S}. It has a box size of $100~\mathrm{cMpc/h}$ and $2\times1792^3$ particles. The resolution elements for dark matter and gas have masses of $1.1\times 10^7~M_{\odot}/h$ and $2.2\times 10^6~M_{\odot}/h$, respectively. The base cosmology corresponds to the results of WMAP7 ~\citep{2011ApJS..192...18K}, i.e., $\Omega_0=0.275$, $\Omega_l=0.725$, $\Omega_b=0.046$, $\sigma_8=0.816$, $h = 0.701$, $n_s=0.968$.  The simulation includes a full modeling of gravity + gas hydrodynamics, as well as a wide range of subgrid recipes for the modeling of star formation ~\citep{2003MNRAS.339..289S}, black hole growth and feedback processes. Haloes were identified using a Friends-of-Friends (FOF) group finder ~\citep{1985ApJ...292..371D}. Within these haloes, self-bound substructures/subhaloes were identified using \texttt{SUBFIND} ~\citep{Springel2001, 2005MNRAS.364.1105S}. Galaxies are identified with the stellar matter components of subhaloes.

For the modeling of black hole growth, a feedback prescription is adopted as detailed in the literature ~\citep{2005Natur.433..604D, 2005MNRAS.361..776S}. In particular, seed black holes of mass $5\times 10^{5}~M_{\odot}/h$ are inserted into haloes of mass $\gtrsim 5\times 10^{10}~M_{\odot}/h$~(if they do not already contain a black hole). Once seeded, black hole growth occurs via gas accretion at a rate given by $\dot{M}_{bh}={4\pi G^2 M_{bh}^2 \rho}/{(c_s^2+v_{bh}^2)^{3/2}}$ where $\rho$ and $c_s$ are the density and sound speed of the ISM gas~(cold phase); $v_{bh}$ is the relative velocity between the black hole and the gas in its vicinity. A radiative efficiency of $10\%$ of the accreted gas is released as radiation. The accretion rate is allowed to be mildly super-Eddington, i.e., limited to two times the Eddington accretion rate. A fraction~($5\%$) of the radiated energy couples to the surrounding gas as black hole~(or AGN) feedback ~\citep{2005Natur.433..604D}.  Note that unlike some previous work, the accretion rate in \mbii\ follows the prescription in~\citet{Pelupessy2007} which does not use any artificial boost factor. For the modeling of black hole mergers, two black holes are considered to be merged if their separation distance is within the spatial resolution of the simulation~(the SPH smoothing length), and their relative speeds are lower than the local sound speed of the medium.

For the galaxy photometry, the SEDs of the host galaxies were first obtained by summing up the contributions from the individual star particles. The stellar SEDs were modelled using the \texttt{PEGASE-2}~\citep{1999astro.ph.12179F} stellar population synthesis code with a Salpeter IMF. The galaxy SEDs are finally convolved with the desired filter function to obtain the broad band photometry~(SDSS $r$-band magnitude). 

Following common practice, the stellar mass is determined within a 3D spherical aperture of 30 kpc as a proxy of the observed stellar mass in the \mbii\ simulation. It has been shown that this definition reproduces a stellar mass function that is consistent with observational measurements~\citep{Pillepich2018}. Furthermore, the stellar mass in this physical aperture provides good agreement to those measured within Petrosian radii in observational studies ~\citep{Schaye2015}. For further details regarding the \mbii\ simulation, we refer the reader to the reference~\citep{2015MNRAS.450.1349K}.

\subsubsection{Semi-analytic model}
\label{sec_intro_SAM}
The Semi-analytic model (\sam) is fully described in~\citet{Menci2016}. Here, we highlight the main points with respect to our study. The merger trees of dark matter haloes are generated through a Monte Carlo procedure by adopting merger rates using an Extended Press \& Schechter formalism~\citep{Lacey1993} assuming a Cold Dark Matter power spectrum of perturbations. For dark matter halos  that merge with a larger halo, we assess the impact of dynamical friction to determine whether it will survive as a satellite, or sink to the centre to increase the mass of the central dominant galaxy; binary interactions (fly-by and merging), among satellite sub-halos, are also described by the model. In each halo, we compute the amount of gas which cools due through atomic processes and settles into a rotationally-supported disk~\citep{Mo1998}. The gas is converted into stars through three different channels: (1) quiescent star formation gradually converting the gas into stars over long timescales $\sim 1$~Gyr, (2) starbursts following galaxy interactions, occurring on timescales $\lesssim 100$~Myr, associated to BH feeding, (3)~internal disc instabilities triggering loss of angular momentum resulting into gas inflows toward the centre thus feeding star formation and BH accretion. The energy released by the supernovae associated with the total star formation returns a fraction of the disc gas into a hot phase (stellar feedback). 

The semi-analytic model includes BH growth from primordial seeds. These are assumed to originate from PopIII stars with a mass $M_{seed}=100\,M_{\odot}$~\citep{Madau2001}, and to be initially present in all galaxy progenitors. We consider two BH feeding modes: accretion triggered by galaxy interactions and internal disc instabilities. These are described in detail in our previous work~\citep{Menci2016}, and briefly described below.\newline
1) BH accretion triggered by interactions. The interaction rate $\tau_r^{-1}=n_T\,\Sigma (r_t,v_c,V)\,V_{rel} (V)$ for galaxies with relative velocity $V_{rel}$ and number density $n_T$ in a common DM halo determines the probability for encounters, 
either fly-by or  merging, through the corresponding cross sections $\Sigma$ given in~\citet{Menci2014}. The fraction of
gas destabilized in each interaction corresponds to the loss $\Delta j$ of orbital angular momentum $j$, and depends on the mass ratio of the merging partners $M'/M$ and on the impact factor $b$. \newline
2) BH accretion induced by disc instabilities. We assume these to arise  in  galaxies with disc mass exceeding~\citep{Efstathiou1982} $M_{crit} =  {v_{max}^2 R_{d}/ G \epsilon}$ with $\epsilon=0.75$, where $v_{max}$ is the maximum circular velocity associated to each halo ~\citep{Mo1998}. 
Such a criterion strongly suppresses the probability for disc instabilities to occur not only in massive, gas-poor galaxies, but also in 
dwarf galaxies characterized by small values of the gas-to-DM mass ratios.
The instabilities induce loss of angular momentum resulting into strong inflows that we compute following the 
description in~\citet{Hopkins2011}, recast and extended as in~\citet{Menci2014}. 

Finally, the \sam\ model includes a detailed treatment of AGN feedback, presented and discussed in~\citet{Menci2008}.
This is assumed to stem from the fast winds with velocity up to
$10^{-1}c$ observed in the central regions of AGNs~\citep{Chartas2002, Pounds2003}.  
These supersonic outflows compress the gas into a blast wave terminated by
a leading shock front, which  moves outwards with a lower but still
supersonic speed and sweeps out the surrounding medium. Eventually,
this medium is expelled from the galaxy. The model follows in detail the expansion of the 
blast wave through the galaxy disc, and computes the fraction of gas expelled from the galaxy.  
These depend on the ratio $\Delta E/E$ between the  energy injected into the galactic gas 
(taken to be proportional to the energy radiated by the 
AGN through the efficiency $\epsilon_{AGN}=5\times 10^{-2}$)
and the thermal energy of the unperturbed gas (see~\citep{Menci2008} for details). 

We note that the absolute determination of stellar mass carries significant uncertainty, both observationally and theoretically. Depending on definitions of stellar mass, on the assumed initial mass function, and possibly on the implementation of star formation in the models, the absolute value of \mstar\ (hence the absolute normalization, i.e., \mbh/\mstar) can vary by up to a factor of two. In contrast, the scatter around the mean correlation is a relative quantity, which is less affected by this uncertainty. Thus, in this work, we mainly focus on the scatter as a diagnostic tool, even though in the future more information could be extracted by this kind of comparison if the 
better measures~(i.e., more consistent with techniques for determining observed \mstar) of stellar masses can be defined for the simulated galaxies.

\section{Comparison Results}
\label{sec:result}
\begin{figure}[t]
\includegraphics[width=1.2\linewidth]{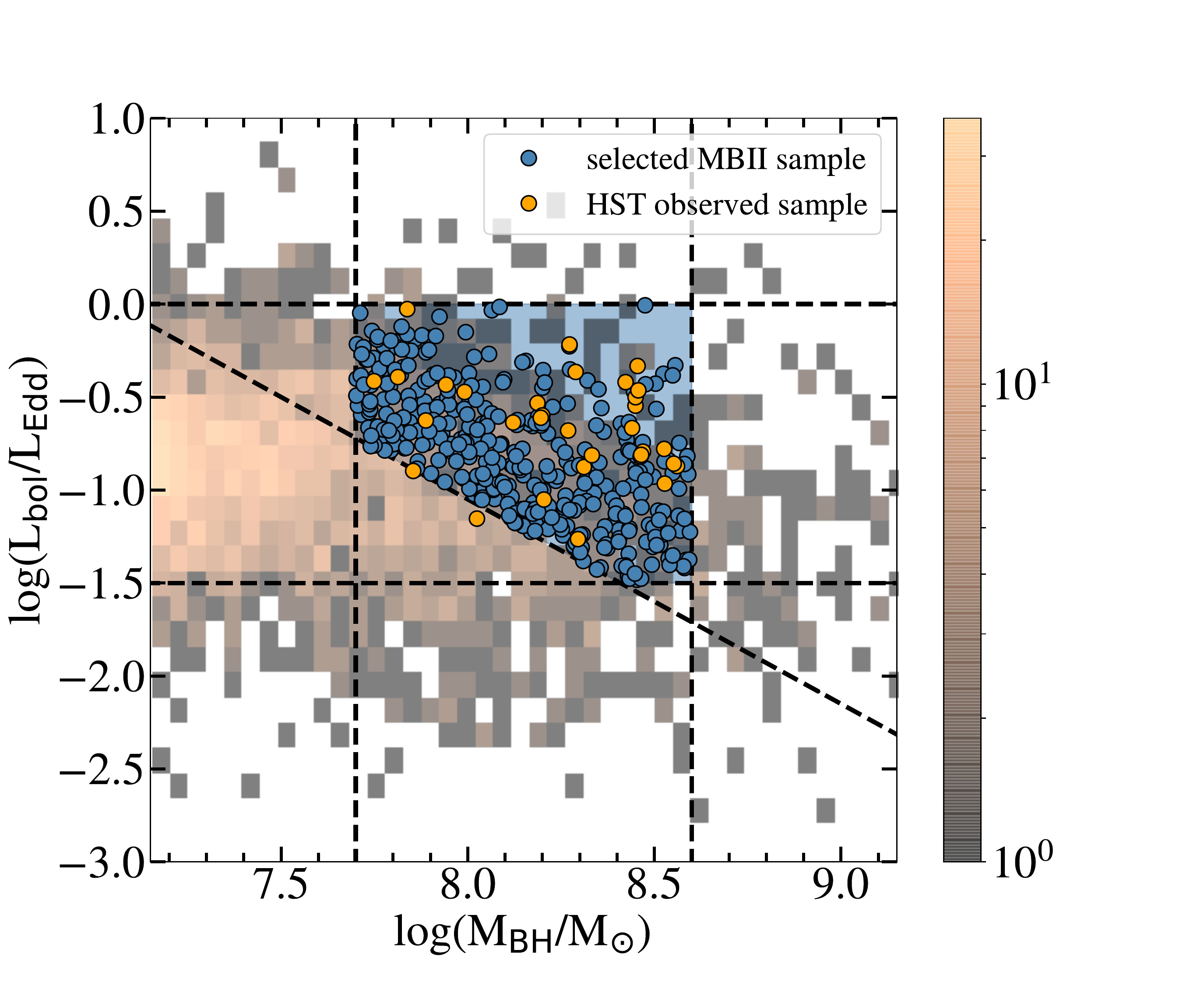}
\caption{Equivalent selection window adopted for the \hst-observed and  \mbii\ simulated samples. The background cloud (in light brown/grey) shows the simulated number density of the overall \mbii\ sample at $z=1.5$. We added random uncertainty to the simulation and select those that fall into the target region (i.e., small blue circles). The small orange circles are the \hst\ observed sample.}
\label{fig:selectfunc}
\end{figure}

Using \mbii, we identify a sample of simulated AGNs at $z=1.5$ and compare their predicted scaling relations to the observed ones. We take the measurement uncertainty and selection biases into account to ensure a fair comparison. First, we inject random noise to the simulated sample to mimic the scatter in our data due to measurement errors, i.e., $\Delta$\mbh $=0.4$~dex, $\Delta$\mr$=0.3$~dex, $\Delta$\mstar$=0.17$~dex, and $\Delta L_{\rm bol}=0.03$~dex, respectively. We then select the sample that falls into the same targeting window to match the observed sample (Figure~\ref{fig:selectfunc}). 

\begin{figure*}[t]
\begin{tabular}{c c}
\includegraphics[trim = 0mm 0mm 65mm 0mm, clip, width=0.47\linewidth]{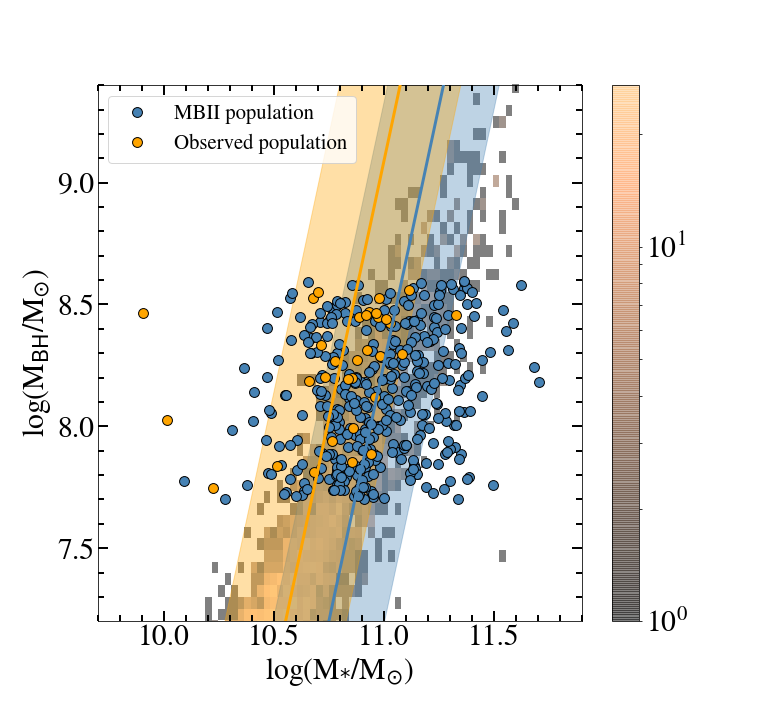} &
\includegraphics[trim = 0mm 0mm 65mm 0mm, clip, width=0.47\linewidth]{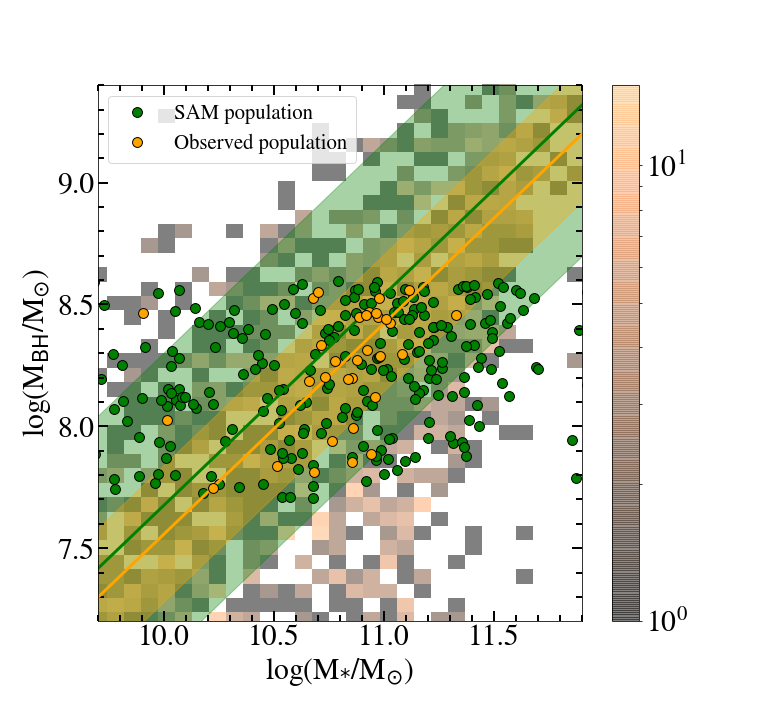} \\
\end{tabular}
\caption{(Left) Comparison of the observed (orange dots) and simulated (blue dots) \mbh--\mstar\ relation. The blue line is the best-fit result for the \mbii\ sample, with the colored region indicating the standard deviation of the residual. The background cloud (in light brown/grey) shows the simulated number density, as Figure~\ref{fig:selectfunc}. By fixing the slope to match the simulated data, the orange color shows the result for the observed data set. The grey cells in the background show the full \mbii\ simulated SMBHs. (Right) The equivalent plot is displayed for the \sam\ sample (green color) in the right panel.}
\label{fig:MM_comp}
\end{figure*}

In the left panel of Figure~\ref{fig:MM_comp}, we compare the relation \mbh--\mstar~between observations and the \mbii\ simulation. It is clear that the simulated and observed samples are in good agreement. 
To quantify the agreement between the simulated and observed data, we use a linear regression to fit their relations. Our selection window has a hard cut on the \mbh\ value (i.e., vertical direction in Figure~\ref{fig:selectfunc}), and thus the scatter on the host properties are larger (horizontal direction). Thus, we fit the host properties (i.e., \mstar) as a function of \mbh. We adopt the {\sc Scipy} package to estimate the best-fit inference for the simulated sample. We then fit the observations based on the same slope value. The comparison results are shown in Figure~\ref{fig:MM_comp} (left panel), with the standard deviation of the residual indicated by the colored regions.
To estimate the {\it observed} scatter of the sample, we calculate the standard deviation of the fitted residual based on \mstar\ (i.e., along the $x$-axis). The histogram of the residual is presented in Figure~\ref{fig:offset_comp} (left panel). We find the standard deviation of the residual for observed and \mbii\ sample are similar, i.e., both equal to $\sim0.3$~dex. We manually change the slope value by its uncertainty level and find that the corresponding {\it observed} scatter barely changes ($<1\%$), meaning that the inferred scatter weakly depends on the fixed slope. To understand how much of the scatter derives from random noise, we measure the scatter of \mbii\ without injecting any noise in the data. Adopting the same selecting window and using the linear regression approach, we estimate the scatter value as $\sim0.1$~dex. Note that this $\sim0.1$~dex scatter level is also controlled by our selection window, and thus it should not be considered as the intrinsic scatter of the overall sample. In this particular work, we simply apply a common standard method (i.e., consideration of the measured residual in \mstar) to achieve a direct comparison between different samples. We perform the Kolmogorov-Smirnov (KS) test of the scatter distribution between observed and \mbii\ sample and infer the p-value as $\sim0.1$. Considering that the simulation sample has been processed to have the same uncertainty level and selection effect, we expect the \mbii\ sample and the observational sample have the same intrinsic scatter level. We adopt the python package {\sc Linmix}~\citep{Kelly2007} to estimate the {\it intrinsic} scatter based on the \mbii\ overall sample and obtain a level of $0.25$~dex.


We also compare our data to predictions by the \sam. In contrast to N-body simulations which produce individual objects, the \sam\ uses an interaction-driven model~\citep{Menci2014} to calculate the number density of the galaxies. To make a direct comparison, we first randomly produce an overall sample based on the \sam\ predicted number density at $z=1.5$. Then, as in the \mbii\ analysis, we inject random scatter in the sample to account for uncertainties and apply the observational selection function. The resulting comparison of the  \mbh-\mstar\ relation is shown in Figure~\ref{fig:MM_comp} (right panel). We find that the best-fit result by the \sam\ model is well matched to the observation. However, the scatter of the \sam\ model is significantly larger ($\sim0.7$~dex) than observed (this is the total scatter accounting for the intrinsic scatter in the \sam\ distribution, observational uncertainties, and selection effects). Even without injecting random noise to \sam\ data, we find that the scatter (considering selection effects) would be $\sim0.6$~dex.

\begin{figure*}[t]
\begin{tabular}{c c}
\includegraphics[trim = 0mm 0mm 61mm 0mm, clip, width=0.47\linewidth]{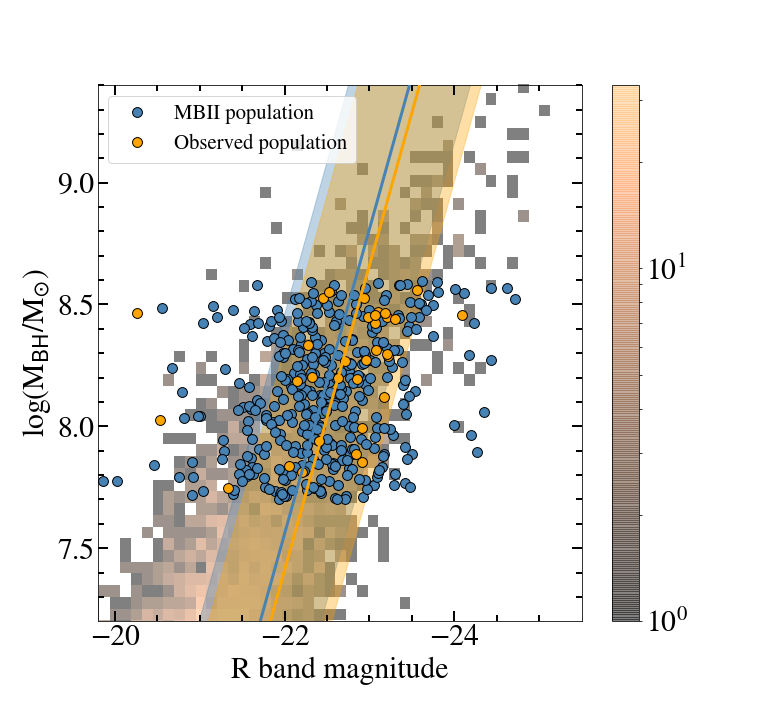} &
\includegraphics[trim = 0mm 0mm 61mm 0mm, clip, width=0.47\linewidth]{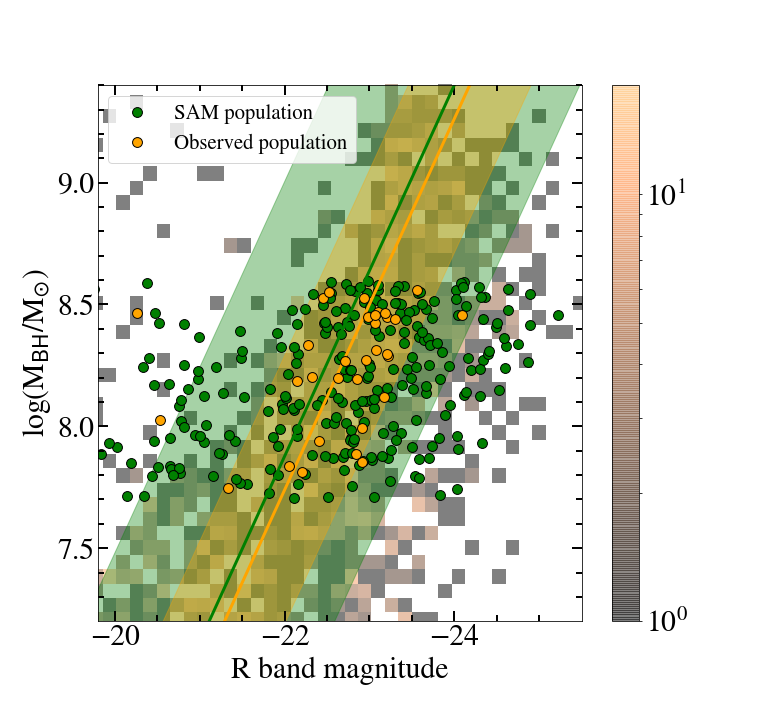} \\
\end{tabular}
\caption{Same as the Figure~\ref{fig:MM_comp}, but for \mbh-\mr\ relation.}
\label{fig:ML_comp}
\end{figure*}
\begin{figure*}[t]
\begin{tabular}{c c}
\includegraphics[width=0.45\linewidth]{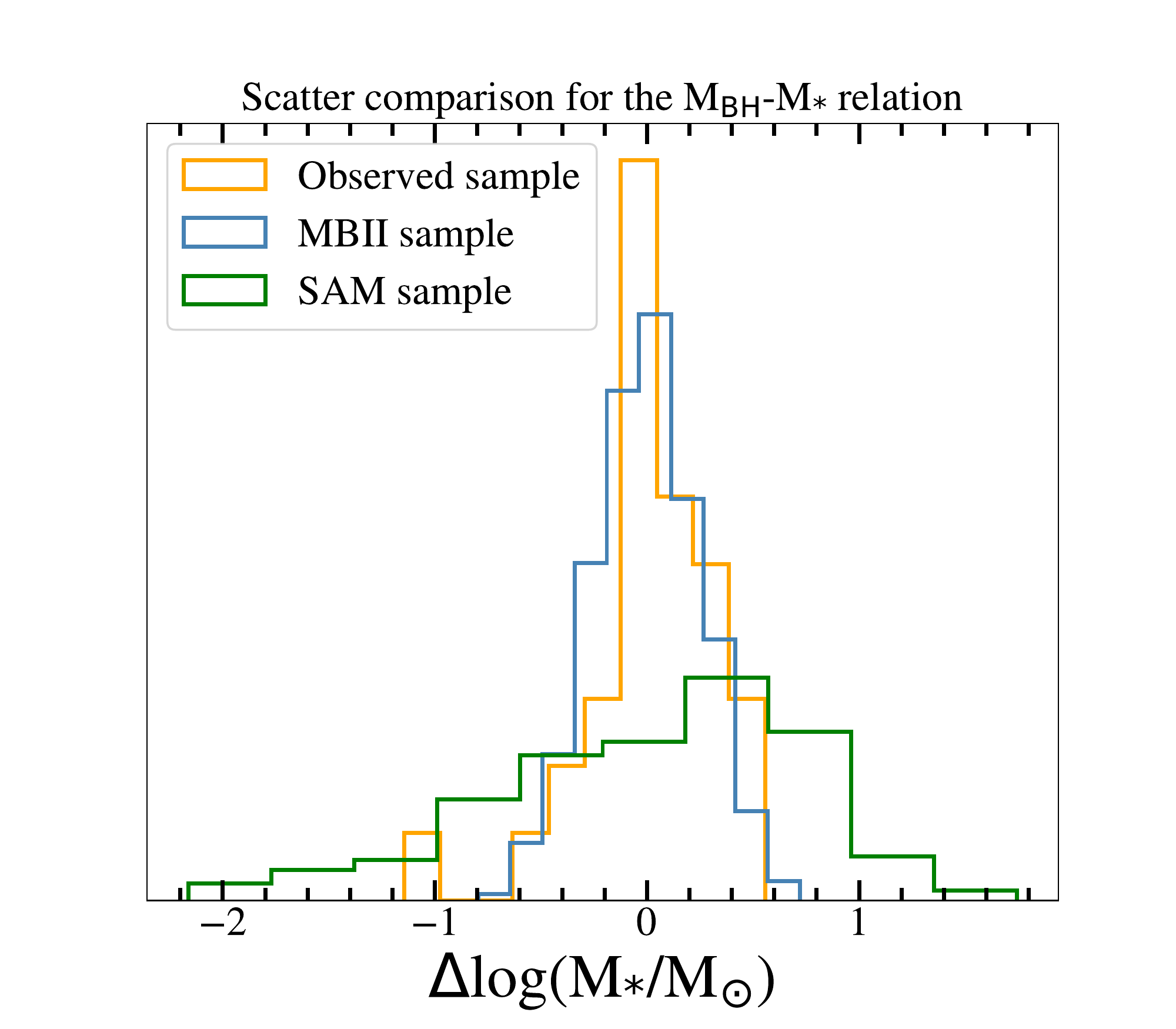} &
\includegraphics[width=0.45\linewidth]{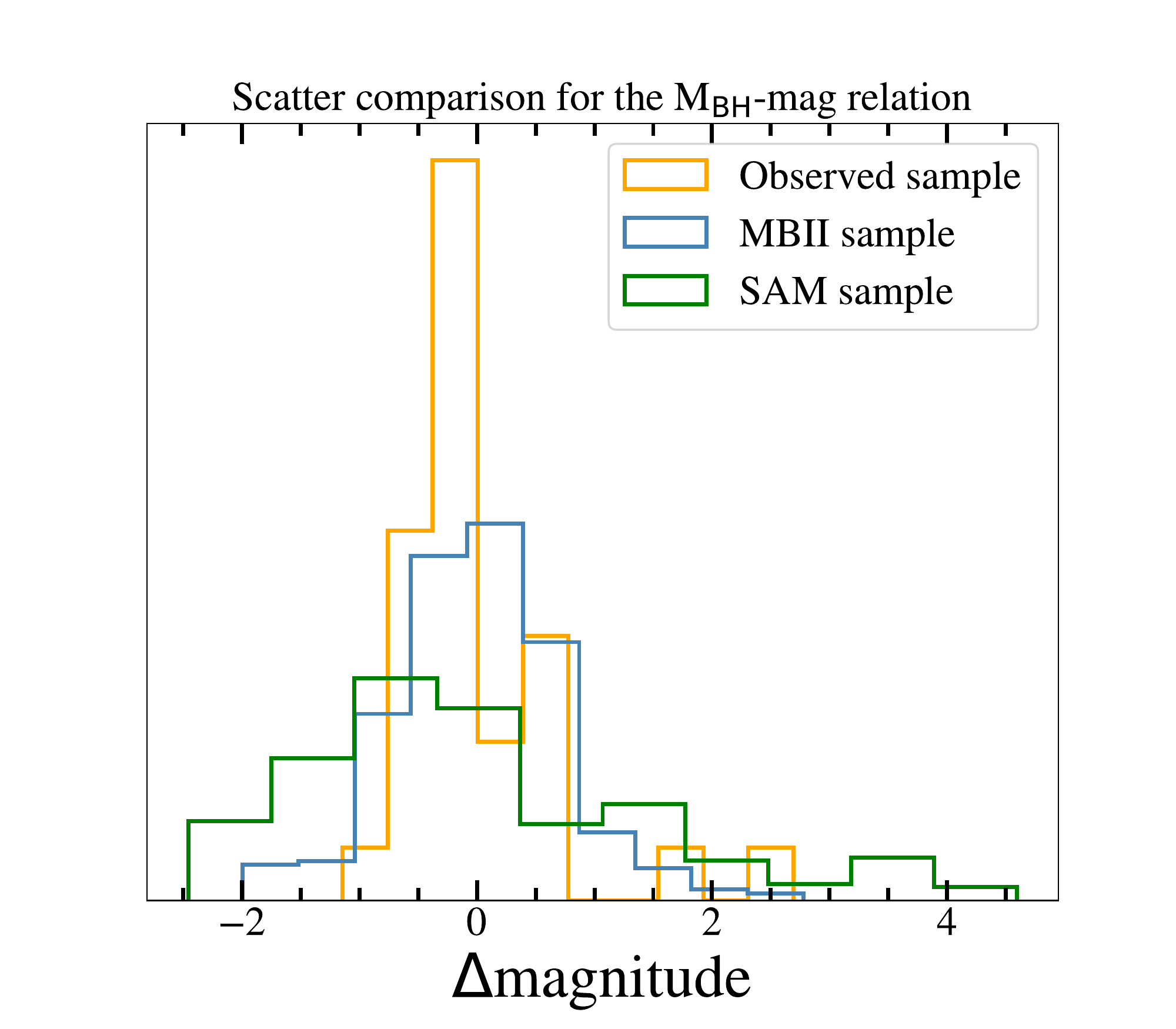} \\
\end{tabular}
\caption{The histogram of the scatter (i.e., residuals in the linear relation). The standard deviation for these distribution are $\sim0.3$~dex, $\sim0.3$~dex and $\sim0.7$~dex for observed sample, \mbii\ sample and \sam\ sample, respectively, for both \mbh-\mstar\ and \mbh-\mr\ relations.
}
\label{fig:offset_comp}
\end{figure*}

We also present the comparison of the \mbh-\mr\ relations in Figure~\ref{fig:ML_comp} and the comparison of the scatter in Figure~\ref{fig:offset_comp} (right panel). The results are similar to \mbh-\mstar\ relations.

\begin{figure}[t]
\includegraphics[width=0.9\linewidth]{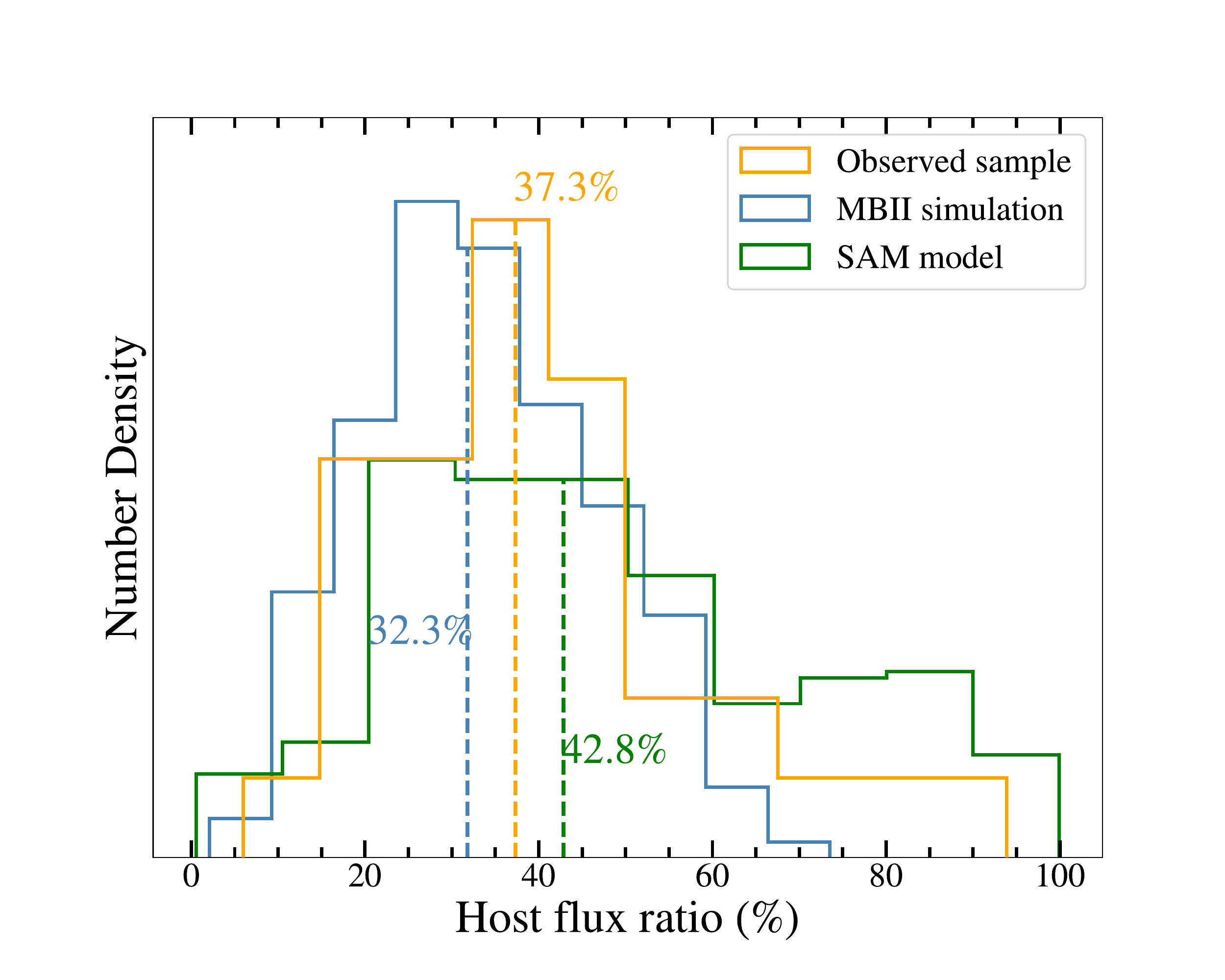}
\caption{Comparison of the host-to-total flux ratio distribution for the three referred samples, median value indicated. The observational selection function is applied to the simulated samples to allow for a fair comparison.
}
\label{fig:comp_hist}
\end{figure}

To test if any unexpected selection effects exist, we compare the distribution of the host-to-total flux ratio among these three samples. For the observed sample, we calculate the flux ratio in the \hst/WFC3 band. For the simulated sample, we consider the AGN bolometric correction~\citep{Elvis1994} to estimate the AGN flux in the WFC3/F125W band. We compare their host-total flux distribution in Figure~\ref{fig:comp_hist} and find that the three samples are well matched to each other. The median values for the flux ratio distribution of the observed, \mbii, \sam\ sample are $37.3\%$, $32.3\%$, and $42.8\%$, respectively. We perform the KS test the inferred p-values are $0.34$ (for observed -- \mbii) and $0.14$ (for observed -- \sam), respectively. These results indicate that one cannot reject the hypothesis that the distributions of the three samples are the same at $10\%$.

\section{Concluding Remarks}
\label{sec:conclusion}
In terms of their central distribution, the \mbii\ simulation and \sam\ model are both in agreement with the observational data. However, the scatter indicates a difference between the two models. The  the scatter in \mbii\ simulation is consistent with that in the observations ($\sim0.3$~dex), while \sam\ sample has a significantly larger amount of scatter ($\sim0.7$~dex). Thus, the implementation of AGN feedback in \mbii\ passes our stringent observational test. The \sam\ model also includes the AGN feedback. However, in contrast to \mbii, the feeding process for SMBH accretion is driven by the 2-body interaction of galaxy mergers, and thus may be more stochastic and lead to larger scatter. More specifically, in the \sam\ model the encounters are assumed to trigger the feedback, and the fraction of gas that feeds the SMBH is related to the parameters of the encounter, which introduces additional scatter since it depends on the properties of both interacting galaxies (see Section~\ref{sec_intro_SAM}). As a result, the \sam\ cloud extends to the high \mbh\ with low stellar mass, which may not exist. The consistency between \mbii\ and observations provides important observational evidence in support of the hypothesis that there is a causal link (i.e., AGN feedback) between the evolution of SMBH and that of their host galaxies. It may also indicate that mergers do not play a dominate role in fueling SMBHs as supported by many observational studies \citep{Ellison2011, Silverman2011,Mechtley2016,Goulding2018}.

The fact that one model (i.e., \sam) does not agree with the observed dispersion is not direct evidence that supports all the physical assumptions implemented in the other model (i.e., \mbii), especially when these two models adopt completely different numerical techniques. At present, it is still unclear how much difference in the observed scatter between \mbii\ and \sam\ is due to the different recipes of triggering, hierarchical merging, gas fueling, and AGN feedback. For instance, even without introducing AGN feedback, the predictions by \citet{2017MNRAS.464.2840A} - based on torques owing to disc instabilities as 
 drivers for black hole feeding - could also have low scatter products in their simulation.
This may indicate that the origin of the smaller scatter in the N-body simulations is related to fact that the considered feeding mechanisms (Bondi accretion or disc instabilities) depend only on the properties of the black hole and host galaxy. 
This strongly differs from the \sam\ assumption that two-body process (interactions) are the main trigger for black hole accretion. 
As for the role of feedback, it will be more insightful to carry out a comparative test based on one numerical model and altering the AGN feedback prescription, while fixing all other conditions \citep[see][]{Hopkins2009}.

It is not straightforward to implementing new physical assumptions to \sam\  in order to solve the tension as discovered in the scatter. 
Increasing, for example, the efficiency of AGN feedback would change the colors of massive galaxies, while changing the Supernovae feedback efficiency would result in a different slope of the galaxy luminosity functions at the faint end, which are constrained by other data. Most importantly, these changes would not appreciably affect the scatter, which originates from the assumption of interactions as triggers for BH accretion. In the \sam\ model, it is possible to switch to disk instabilities as triggers for accretion. However, this channel alone would not be able to yield the accretion rates necessary to power the most luminous quasars~\citep[see][]{Menci2014}. Implementing ``ad-hoc'', i.e., completely phenomenological and parametrized laws for the accretion, would require a long and detailed exploration of the possible parameters and the impact of each choice not only on the observables connected to AGN, but also on the properties of the galaxy populations (e.g., colors, luminosity functions, etc.). In addition, such a new approach with massive efforts would possibly not provide deeper insights on the physical mechanisms driving the AGN-galaxy connection.

Without a direct physical mechanism, it has been shown that, due to the central limit theorem~\citep{Peng2007, Jahnke2011, Hirschmann2010}, scaling relations may emerge from random mergers, starting from a stochastic cloud in the early universe. Under this scenario, the scatter of the scaling relations has to increase with redshift. Our observations contradict this hypothesis. In fact, the inferred {\it intrinsic} scatter of our observed sample (i.e., $\sim0.25$~dex) is even no more significant than the typical scatter of local relations reported in the literature ~\citep[][i.e., $\gtrsim0.35$~dex]{Kormendy13, Gul++09, Reines2015}. Of course, the intrinsic scatter of our high-$z$ sample could be inaccurate, since we use the \mbii\ overall sample as a proxy to estimate the level of intrinsic scatter for the real data, but it is unlikely that systematic errors would conspire to {\it reduce} scatter.
Also, the observed \mbh\ are estimated using the robust \halpha\ line, which could have lower uncertainties level than expected (i.e., $\Delta$\mbh$<0.4$~dex), resulting in an overestimating of the error-budget and thus underestimating of intrinsic scatter, but again it is hard to imagine that the \halpha\ \mbh\ estimators be much more precise than an factor of two.

Our sample of 32 AGN systems covers the range $1.2<z<1.7$. In principle it would be interesting to consider the evolution trend of the mass relation within the redshift range and make comparisons with the simulation as a function of cosmic time. Unfortunately, the limited sample size and precision of the measurement are not sufficient to resolve the evolution within this redshift range (see Figure~8 in D20). Thus, we only consider the sample with a single redshift bin at $z\sim1.5$ to compare with the simulations and the local measurements.

Extending this study to even higher redshift would be very beneficial, probing closer to the epoch of formation of massive galaxies and SMBHs. For higher redshift, the {\it James Webb Space Telescope} may provide high-quality imaging data of AGNs at redshift up to $z\sim7$. In the low redshift Universe, wide-area surveys with Subaru/HSC, LSST, and WFIRST offer much promise to build samples for studying these mass ratios and dependencies on other factors (e.g., environment).





\acknowledgments
Based in part on observations made with the NASA/ESA Hubble Space Telescope, obtained at the Space Telescope Science Institute, which is operated by the Association of Universities for Research in Astronomy, Inc., under NASA contract NAS 5-26555. These observations are associated with programs \#15115. Support for this work was provided by NASA through grant number HST-GO-15115 from the Space Telescope Science Institute, which is operated by AURA, Inc., under NASA contract NAS 5-26555. The authors fully appreciate input from Simon Birrer, Matthew A. Malkan. XD and TT acknowledge support by the Packard Foundation through a Packard Research fellowship to TT. JS is supported by JSPS KAKENHI Grant Number JP18H01251 and the World Premier International Research Center Initiative (WPI), MEXT, Japan. TDM acknowledges funding from NSF ACI-1614853,  NSF AST-1616168 NASA ATP 80NSSC18K101 and NASA ATP 17-0123. We acknowledge support from INAF under PRIN SKA/ CTA FORECaST and PRIN SKA-CTA-INAF ASTRI/CTA Data Challenge. 

\bibliography{reference}

\begin{thebibliography}{}
\expandafter\ifx\csname natexlab\endcsname\relax\def\natexlab#1{#1}\fi
\providecommand{\url}[1]{\href{#1}{#1}}
\providecommand{\dodoi}[1]{doi:~\href{http://doi.org/#1}{\nolinkurl{#1}}}
\providecommand{\doeprint}[1]{\href{http://ascl.net/#1}{\nolinkurl{http://ascl.net/#1}}}
\providecommand{\doarXiv}[1]{\href{https://arxiv.org/abs/#1}{\nolinkurl{https://arxiv.org/abs/#1}}}

\bibitem[{{Angl{\'e}s-Alc{\'a}zar} {et~al.}(2017){Angl{\'e}s-Alc{\'a}zar},
  {Dav{\'e}}, {Faucher-Gigu{\`e}re}, {{\"O}zel}, \&
  {Hopkins}}]{2017MNRAS.464.2840A}
{Angl{\'e}s-Alc{\'a}zar}, D., {Dav{\'e}}, R., {Faucher-Gigu{\`e}re}, C.-A.,
  {{\"O}zel}, F., \& {Hopkins}, P.~F. 2017, \mnras, 464, 2840,
  \dodoi{10.1093/mnras/stw2565}

\bibitem[{{Bennert} {et~al.}(2011){Bennert}, {Auger}, {Treu}, {Woo}, \&
  {Malkan}}]{Bennert11}
{Bennert}, V.~N., {Auger}, M.~W., {Treu}, T., {Woo}, J.-H., \& {Malkan}, M.~A.
  2011, \apj, 742, 107, \dodoi{10.1088/0004-637X/742/2/107}

\bibitem[{{Bhowmick} {et~al.}(2019){Bhowmick}, {DiMatteo}, {Eftekharzadeh}, \&
  {Myers}}]{Bhowmick2019}
{Bhowmick}, A.~K., {DiMatteo}, T., {Eftekharzadeh}, S., \& {Myers}, A.~D. 2019,
  \mnras, 485, 2026, \dodoi{10.1093/mnras/stz519}

\bibitem[{{Birrer} \& {Amara}(2018)}]{lenstronomy}
{Birrer}, S., \& {Amara}, A. 2018, Physics of the Dark Universe, 22, 189,
  \dodoi{10.1016/j.dark.2018.11.002}

\bibitem[{{Cen}(2015)}]{Cen2015}
{Cen}, R. 2015, \apjl, 805, L9, \dodoi{10.1088/2041-8205/805/1/L9}

\bibitem[{{Chartas} {et~al.}(2002){Chartas}, {Brandt}, {Gallagher}, \&
  {Garmire}}]{Chartas2002}
{Chartas}, G., {Brandt}, W.~N., {Gallagher}, S.~C., \& {Garmire}, G.~P. 2002,
  \apj, 579, 169, \dodoi{10.1086/342744}

\bibitem[{{Civano} {et~al.}(2016){Civano}, {Marchesi}, {Comastri}, {Urry},
  {Elvis}, {Cappelluti}, {Puccetti}, {Brusa}, {Zamorani}, {Hasinger},
  {Aldcroft}, {Alexander}, {Allevato}, {Brunner}, {Capak}, {Finoguenov},
  {Fiore}, {Fruscione}, {Gilli}, {Glotfelty}, {Griffiths}, {Hao}, {Harrison},
  {Jahnke}, {Kartaltepe}, {Karim}, {LaMassa}, {Lanzuisi}, {Miyaji}, {Ranalli},
  {Salvato}, {Sargent}, {Scoville}, {Schawinski}, {Schinnerer}, {Silverman},
  {Smolcic}, {Stern}, {Toft}, {Trakhtenbrot}, {Treister}, \&
  {Vignali}}]{Civano2016}
{Civano}, F., {Marchesi}, S., {Comastri}, A., {et~al.} 2016, \apj, 819, 62,
  \dodoi{10.3847/0004-637X/819/1/62}

\bibitem[{{Davis} {et~al.}(1985){Davis}, {Efstathiou}, {Frenk}, \&
  {White}}]{1985ApJ...292..371D}
{Davis}, M., {Efstathiou}, G., {Frenk}, C.~S., \& {White}, S.~D.~M. 1985, \apj,
  292, 371, \dodoi{10.1086/163168}

\bibitem[{{DeGraf} {et~al.}(2015){DeGraf}, {Di Matteo}, {Treu}, {Feng}, {Woo},
  \& {Park}}]{DeG++15}
{DeGraf}, C., {Di Matteo}, T., {Treu}, T., {et~al.} 2015, \mnras, 454, 913,
  \dodoi{10.1093/mnras/stv2002}

\bibitem[{{Di Matteo} {et~al.}(2005){Di Matteo}, {Springel}, \&
  {Hernquist}}]{2005Natur.433..604D}
{Di Matteo}, T., {Springel}, V., \& {Hernquist}, L. 2005, \nat, 433, 604,
  \dodoi{10.1038/nature03335}

\bibitem[{{Ding} {et~al.}(2020){Ding}, {Silverman}, {Treu}, {Schulze},
  {Schramm}, {Birrer}, {Park}, {Jahnke}, {Bennert}, {Kartaltepe}, {Koekemoer},
  {Malkan}, \& {Sanders}}]{Ding2020}
{Ding}, X., {Silverman}, J., {Treu}, T., {et~al.} 2020, \apj, 888, 37,
  \dodoi{10.3847/1538-4357/ab5b90}

\bibitem[{{Efstathiou} {et~al.}(1982){Efstathiou}, {Lake}, \&
  {Negroponte}}]{Efstathiou1982}
{Efstathiou}, G., {Lake}, G., \& {Negroponte}, J. 1982, \mnras, 199, 1069,
  \dodoi{10.1093/mnras/199.4.1069}

\bibitem[{{Ellison} {et~al.}(2011){Ellison}, {Patton}, {Mendel}, \&
  {Scudder}}]{Ellison2011}
{Ellison}, S.~L., {Patton}, D.~R., {Mendel}, J.~T., \& {Scudder}, J.~M. 2011,
  \mnras, 418, 2043, \dodoi{10.1111/j.1365-2966.2011.19624.x}

\bibitem[{{Elvis} {et~al.}(1994){Elvis}, {Wilkes}, {McDowell}, {Green},
  {Bechtold}, {Willner}, {Oey}, {Polomski}, \& {Cutri}}]{Elvis1994}
{Elvis}, M., {Wilkes}, B.~J., {McDowell}, J.~C., {et~al.} 1994, \apjs, 95, 1,
  \dodoi{10.1086/192093}

\bibitem[{{Ferrarese} \& {Merritt}(2000)}]{F+M00}
{Ferrarese}, L., \& {Merritt}, D. 2000, \apjl, 539, L9, \dodoi{10.1086/312838}

\bibitem[{{Fioc} \& {Rocca-Volmerange}(1999)}]{1999astro.ph.12179F}
{Fioc}, M., \& {Rocca-Volmerange}, B. 1999, arXiv e-prints, astro.
\newblock \doarXiv{astro-ph/9912179}

\bibitem[{{Goulding} {et~al.}(2018){Goulding}, {Greene}, {Bezanson}, {Greco},
  {Johnson}, {Leauthaud}, {Matsuoka}, {Medezinski}, \&
  {Price-Whelan}}]{Goulding2018}
{Goulding}, A.~D., {Greene}, J.~E., {Bezanson}, R., {et~al.} 2018, \pasj, 70,
  S37, \dodoi{10.1093/pasj/psx135}

\bibitem[{{G{\"u}ltekin} {et~al.}(2009){G{\"u}ltekin}, {Richstone}, {Gebhardt},
  {Lauer}, {Tremaine}, {Aller}, {Bender}, {Dressler}, {Faber}, {Filippenko},
  {Green}, {Ho}, {Kormendy}, {Magorrian}, {Pinkney}, \& {Siopis}}]{Gul++09}
{G{\"u}ltekin}, K., {Richstone}, D.~O., {Gebhardt}, K., {et~al.} 2009, \apj,
  698, 198, \dodoi{10.1088/0004-637X/698/1/198}

\bibitem[{{H{\"a}ring} \& {Rix}(2004)}]{H+R04}
{H{\"a}ring}, N., \& {Rix}, H.-W. 2004, \apjl, 604, L89, \dodoi{10.1086/383567}

\bibitem[{{Hirschmann} {et~al.}(2010){Hirschmann}, {Khochfar}, {Burkert},
  {Naab}, {Genel}, \& {Somerville}}]{Hirschmann2010}
{Hirschmann}, M., {Khochfar}, S., {Burkert}, A., {et~al.} 2010, \mnras, 407,
  1016, \dodoi{10.1111/j.1365-2966.2010.17006.x}

\bibitem[{{Hopkins} {et~al.}(2009){Hopkins}, {Murray}, \&
  {Thompson}}]{Hopkins2009}
{Hopkins}, P.~F., {Murray}, N., \& {Thompson}, T.~A. 2009, \mnras, 398, 303,
  \dodoi{10.1111/j.1365-2966.2009.15132.x}

\bibitem[{{Hopkins} \& {Quataert}(2011)}]{Hopkins2011}
{Hopkins}, P.~F., \& {Quataert}, E. 2011, \mnras, 415, 1027,
  \dodoi{10.1111/j.1365-2966.2011.18542.x}

\bibitem[{{Huang} {et~al.}(2018){Huang}, {Di Matteo}, {Bhowmick}, {Feng}, \&
  {Ma}}]{Huang2018}
{Huang}, K.-W., {Di Matteo}, T., {Bhowmick}, A.~K., {Feng}, Y., \& {Ma}, C.-P.
  2018, \mnras, 478, 5063, \dodoi{10.1093/mnras/sty1329}

\bibitem[{{Jahnke} \& {Macci{\`o}}(2011)}]{Jahnke2011}
{Jahnke}, K., \& {Macci{\`o}}, A.~V. 2011, \apj, 734, 92,
  \dodoi{10.1088/0004-637X/734/2/92}

\bibitem[{{Jahnke} {et~al.}(2009){Jahnke}, {Bongiorno}, {Brusa}, {Capak},
  {Cappelluti}, {Cisternas}, {Civano}, {Colbert}, {Comastri}, {Elvis},
  {Hasinger}, {Ilbert}, {Impey}, {Inskip}, {Koekemoer}, {Lilly}, {Maier},
  {Merloni}, {Riechers}, {Salvato}, {Schinnerer}, {Scoville}, {Silverman},
  {Taniguchi}, {Trump}, \& {Yan}}]{Jahnke2009}
{Jahnke}, K., {Bongiorno}, A., {Brusa}, M., {et~al.} 2009, \apjl, 706, L215,
  \dodoi{10.1088/0004-637X/706/2/L215}

\bibitem[{{Kelly}(2007)}]{Kelly2007}
{Kelly}, B.~C. 2007, \apj, 665, 1489, \dodoi{10.1086/519947}

\bibitem[{{Khandai} {et~al.}(2015{\natexlab{a}}){Khandai}, {Di Matteo},
  {Croft}, {Wilkins}, {Feng}, {Tucker}, {DeGraf}, \& {Liu}}]{Khandai2015}
{Khandai}, N., {Di Matteo}, T., {Croft}, R., {et~al.} 2015{\natexlab{a}},
  \mnras, 450, 1349, \dodoi{10.1093/mnras/stv627}

\bibitem[{{Khandai} {et~al.}(2015{\natexlab{b}}){Khandai}, {Di Matteo},
  {Croft}, {Wilkins}, {Feng}, {Tucker}, {DeGraf}, \&
  {Liu}}]{2015MNRAS.450.1349K}
---. 2015{\natexlab{b}}, \mnras, 450, 1349, \dodoi{10.1093/mnras/stv627}

\bibitem[{{Komatsu} {et~al.}(2011){Komatsu}, {Smith}, {Dunkley}, {Bennett},
  {Gold}, {Hinshaw}, {Jarosik}, {Larson}, {Nolta}, {Page}, {Spergel},
  {Halpern}, {Hill}, {Kogut}, {Limon}, {Meyer}, {Odegard}, {Tucker}, {Weiland},
  {Wollack}, \& {Wright}}]{2011ApJS..192...18K}
{Komatsu}, E., {Smith}, K.~M., {Dunkley}, J., {et~al.} 2011, \apjs, 192, 18,
  \dodoi{10.1088/0067-0049/192/2/18}

\bibitem[{{Kormendy} \& {Ho}(2013)}]{Kormendy13}
{Kormendy}, J., \& {Ho}, L.~C. 2013, \araa, 51, 511,
  \dodoi{10.1146/annurev-astro-082708-101811}

\bibitem[{{Lacey} \& {Cole}(1993)}]{Lacey1993}
{Lacey}, C., \& {Cole}, S. 1993, \mnras, 262, 627,
  \dodoi{10.1093/mnras/262.3.627}

\bibitem[{{Lauer} {et~al.}(2007){Lauer}, {Tremaine}, {Richstone}, \&
  {Faber}}]{Lauer2007}
{Lauer}, T.~R., {Tremaine}, S., {Richstone}, D., \& {Faber}, S.~M. 2007, \apj,
  670, 249, \dodoi{10.1086/522083}

\bibitem[{{Lehmer} {et~al.}(2005){Lehmer}, {Brandt}, {Alexander}, {Bauer},
  {Schneider}, {Tozzi}, {Bergeron}, {Garmire}, {Giacconi}, {Gilli}, {Hasinger},
  {Hornschemeier}, {Koekemoer}, {Mainieri}, {Miyaji}, {Nonino}, {Rosati},
  {Silverman}, {Szokoly}, \& {Vignali}}]{Lehmer2005}
{Lehmer}, B.~D., {Brandt}, W.~N., {Alexander}, D.~M., {et~al.} 2005, \apjs,
  161, 21, \dodoi{10.1086/444590}

\bibitem[{{Li} {et~al.}(2019){Li}, {Habouzit}, {Genel}, {Somerville},
  {Terrazas}, {Bell}, {Pillepich}, {Nelson}, {Weinberger}, {Rodriguez-Gomez},
  {Ma}, {Pakmor}, {Hernquist}, \& {Vogelsberger}}]{Li2019}
{Li}, Y., {Habouzit}, M., {Genel}, S., {et~al.} 2019, arXiv e-prints,
  arXiv:1910.00017.
\newblock \doarXiv{1910.00017}

\bibitem[{{Madau} \& {Rees}(2001)}]{Madau2001}
{Madau}, P., \& {Rees}, M.~J. 2001, \apjl, 551, L27, \dodoi{10.1086/319848}

\bibitem[{{Magorrian} {et~al.}(1998){Magorrian}, {Tremaine}, {Richstone},
  {Bender}, {Bower}, {Dressler}, {Faber}, {Gebhardt}, {Green}, {Grillmair},
  {Kormendy}, \& {Lauer}}]{Mag++98}
{Magorrian}, J., {Tremaine}, S., {Richstone}, D., {et~al.} 1998, \aj, 115,
  2285, \dodoi{10.1086/300353}

\bibitem[{{Marconi} \& {Hunt}(2003)}]{M+H03}
{Marconi}, A., \& {Hunt}, L.~K. 2003, \apjl, 589, L21, \dodoi{10.1086/375804}

\bibitem[{{Mechtley} {et~al.}(2016){Mechtley}, {Jahnke}, {Windhorst}, {Andrae},
  {Cisternas}, {Cohen}, {Hewlett}, {Koekemoer}, {Schramm}, \&
  {Schulze}}]{Mechtley2016}
{Mechtley}, M., {Jahnke}, K., {Windhorst}, R.~A., {et~al.} 2016, \apj, 830,
  156, \dodoi{10.3847/0004-637X/830/2/156}

\bibitem[{{Menci} {et~al.}(2016){Menci}, {Fiore}, {Bongiorno}, \&
  {Lamastra}}]{Menci2016}
{Menci}, N., {Fiore}, F., {Bongiorno}, A., \& {Lamastra}, A. 2016, \aap, 594,
  A99, \dodoi{10.1051/0004-6361/201628415}

\bibitem[{{Menci} {et~al.}(2008){Menci}, {Fiore}, {Puccetti}, \&
  {Cavaliere}}]{Menci2008}
{Menci}, N., {Fiore}, F., {Puccetti}, S., \& {Cavaliere}, A. 2008, \apj, 686,
  219, \dodoi{10.1086/591438}

\bibitem[{{Menci} {et~al.}(2014){Menci}, {Gatti}, {Fiore}, \&
  {Lamastra}}]{Menci2014}
{Menci}, N., {Gatti}, M., {Fiore}, F., \& {Lamastra}, A. 2014, \aap, 569, A37,
  \dodoi{10.1051/0004-6361/201424217}

\bibitem[{{Mo} {et~al.}(1998){Mo}, {Mao}, \& {White}}]{Mo1998}
{Mo}, H.~J., {Mao}, S., \& {White}, S.~D.~M. 1998, \mnras, 295, 319,
  \dodoi{10.1046/j.1365-8711.1998.01227.x}

\bibitem[{{Park} {et~al.}(2015){Park}, {Woo}, {Bennert}, {Treu}, {Auger}, \&
  {Malkan}}]{Park15}
{Park}, D., {Woo}, J.-H., {Bennert}, V.~N., {et~al.} 2015, \apj, 799, 164,
  \dodoi{10.1088/0004-637X/799/2/164}

\bibitem[{{Pelupessy} {et~al.}(2007){Pelupessy}, {Di Matteo}, \&
  {Ciardi}}]{Pelupessy2007}
{Pelupessy}, F.~I., {Di Matteo}, T., \& {Ciardi}, B. 2007, \apj, 665, 107,
  \dodoi{10.1086/519235}

\bibitem[{{Peng}(2007)}]{Peng2007}
{Peng}, C.~Y. 2007, \apj, 671, 1098, \dodoi{10.1086/522774}

\bibitem[{{Peng} {et~al.}(2006){Peng}, {Impey}, {Ho}, {Barton}, \&
  {Rix}}]{Peng2006a}
{Peng}, C.~Y., {Impey}, C.~D., {Ho}, L.~C., {Barton}, E.~J., \& {Rix}, H.-W.
  2006, \apj, 640, 114, \dodoi{10.1086/499930}

\bibitem[{{Pillepich} {et~al.}(2018){Pillepich}, {Nelson}, {Hernquist},
  {Springel}, {Pakmor}, {Torrey}, {Weinberger}, {Genel}, {Naiman}, {Marinacci},
  \& {Vogelsberger}}]{Pillepich2018}
{Pillepich}, A., {Nelson}, D., {Hernquist}, L., {et~al.} 2018, \mnras, 475,
  648, \dodoi{10.1093/mnras/stx3112}

\bibitem[{{Pounds} {et~al.}(2003){Pounds}, {King}, {Page}, \&
  {O'Brien}}]{Pounds2003}
{Pounds}, K.~A., {King}, A.~R., {Page}, K.~L., \& {O'Brien}, P.~T. 2003,
  \mnras, 346, 1025, \dodoi{10.1111/j.1365-2966.2003.07164.x}

\bibitem[{{Reines} \& {Volonteri}(2015)}]{Reines2015}
{Reines}, A.~E., \& {Volonteri}, M. 2015, \apj, 813, 82,
  \dodoi{10.1088/0004-637X/813/2/82}

\bibitem[{{Schaye} {et~al.}(2015){Schaye}, {Crain}, {Bower}, {Furlong},
  {Schaller}, {Theuns}, {Dalla Vecchia}, {Frenk}, {McCarthy}, {Helly},
  {Jenkins}, {Rosas-Guevara}, {White}, {Baes}, {Booth}, {Camps}, {Navarro},
  {Qu}, {Rahmati}, {Sawala}, {Thomas}, \& {Trayford}}]{Schaye2015}
{Schaye}, J., {Crain}, R.~A., {Bower}, R.~G., {et~al.} 2015, \mnras, 446, 521,
  \dodoi{10.1093/mnras/stu2058}

\bibitem[{{Schramm} \& {Silverman}(2013)}]{Schramm2013}
{Schramm}, M., \& {Silverman}, J.~D. 2013, \apj, 767, 13,
  \dodoi{10.1088/0004-637X/767/1/13}

\bibitem[{{Schulze} \& {Wisotzki}(2014)}]{Schulze2014}
{Schulze}, A., \& {Wisotzki}, L. 2014, \mnras, 438, 3422,
  \dodoi{10.1093/mnras/stt2457}

\bibitem[{{Sijacki} {et~al.}(2015){Sijacki}, {Vogelsberger}, {Genel},
  {Springel}, {Torrey}, {Snyder}, {Nelson}, \& {Hernquist}}]{Sijacki2015}
{Sijacki}, D., {Vogelsberger}, M., {Genel}, S., {et~al.} 2015, \mnras, 452,
  575, \dodoi{10.1093/mnras/stv1340}

\bibitem[{{Silverman} {et~al.}(2011){Silverman}, {Kampczyk}, {Jahnke},
  {Andrae}, {Lilly}, {Elvis}, {Civano}, {Mainieri}, {Vignali}, {Zamorani},
  {Nair}, {Le F{\`e}vre}, {de Ravel}, {Bardelli}, {Bongiorno}, {Bolzonella},
  {Cappi}, {Caputi}, {Carollo}, {Contini}, {Coppa}, {Cucciati}, {de la Torre},
  {Franzetti}, {Garilli}, {Halliday}, {Hasinger}, {Iovino}, {Knobel},
  {Koekemoer}, {Kova{\v{c}}}, {Lamareille}, {Le Borgne}, {Le Brun}, {Maier},
  {Mignoli}, {Pello}, {P{\'e}rez-Montero}, {Ricciardelli}, {Peng}, {Scodeggio},
  {Tanaka}, {Tasca}, {Tresse}, {Vergani}, {Zucca}, {Brusa}, {Cappelluti},
  {Comastri}, {Finoguenov}, {Fu}, {Gilli}, {Hao}, {Ho}, \&
  {Salvato}}]{Silverman2011}
{Silverman}, J.~D., {Kampczyk}, P., {Jahnke}, K., {et~al.} 2011, \apj, 743, 2,
  \dodoi{10.1088/0004-637X/743/1/2}

\bibitem[{{Springel}(2005)}]{2005MNRAS.364.1105S}
{Springel}, V. 2005, \mnras, 364, 1105,
  \dodoi{10.1111/j.1365-2966.2005.09655.x}

\bibitem[{{Springel} {et~al.}(2005){Springel}, {Di Matteo}, \&
  {Hernquist}}]{2005MNRAS.361..776S}
{Springel}, V., {Di Matteo}, T., \& {Hernquist}, L. 2005, \mnras, 361, 776,
  \dodoi{10.1111/j.1365-2966.2005.09238.x}

\bibitem[{{Springel} \& {Hernquist}(2003)}]{2003MNRAS.339..289S}
{Springel}, V., \& {Hernquist}, L. 2003, \mnras, 339, 289,
  \dodoi{10.1046/j.1365-8711.2003.06206.x}

\bibitem[{{Springel} {et~al.}(2001){Springel}, {White}, {Tormen}, \&
  {Kauffmann}}]{Springel2001}
{Springel}, V., {White}, S. D.~M., {Tormen}, G., \& {Kauffmann}, G. 2001,
  \mnras, 328, 726, \dodoi{10.1046/j.1365-8711.2001.04912.x}

\bibitem[{{Steinborn} {et~al.}(2015){Steinborn}, {Dolag}, {Hirschmann},
  {Prieto}, \& {Remus}}]{Steinborn2015}
{Steinborn}, L.~K., {Dolag}, K., {Hirschmann}, M., {Prieto}, M.~A., \& {Remus},
  R.-S. 2015, \mnras, 448, 1504, \dodoi{10.1093/mnras/stv072}

\bibitem[{{Sun} {et~al.}(2015){Sun}, {Trump}, {Brandt}, {Luo}, {Alexander},
  {Jahnke}, {Rosario}, {Wang}, \& {Xue}}]{Sun2015}
{Sun}, M., {Trump}, J.~R., {Brandt}, W.~N., {et~al.} 2015, \apj, 802, 14,
  \dodoi{10.1088/0004-637X/802/1/14}

\bibitem[{{Thomas} {et~al.}(2019){Thomas}, {Dav{\'e}},
  {Angl{\'e}s-Alc{\'a}zar}, \& {Jarvis}}]{Thomas2019}
{Thomas}, N., {Dav{\'e}}, R., {Angl{\'e}s-Alc{\'a}zar}, D., \& {Jarvis}, M.
  2019, \mnras, 487, 5764, \dodoi{10.1093/mnras/stz1703}

\bibitem[{{Treu} {et~al.}(2007){Treu}, {Woo}, {Malkan}, \&
  {Blandford}}]{Tre++07}
{Treu}, T., {Woo}, J.-H., {Malkan}, M.~A., \& {Blandford}, R.~D. 2007, \apj,
  667, 117, \dodoi{10.1086/520633}

\bibitem[{{Ueda} {et~al.}(2008){Ueda}, {Watson}, {Stewart}, {Akiyama},
  {Schwope}, {Lamer}, {Ebrero}, {Carrera}, {Sekiguchi}, {Yamada}, {Simpson},
  {Hasinger}, \& {Mateos}}]{Ueda2008}
{Ueda}, Y., {Watson}, M.~G., {Stewart}, I.~M., {et~al.} 2008, \apjs, 179, 124,
  \dodoi{10.1086/591083}

\bibitem[{{Vogelsberger} {et~al.}(2014){Vogelsberger}, {Genel}, {Springel},
  {Torrey}, {Sijacki}, {Xu}, {Snyder}, {Nelson}, \&
  {Hernquist}}]{Vogelsberger2014}
{Vogelsberger}, M., {Genel}, S., {Springel}, V., {et~al.} 2014, \mnras, 444,
  1518, \dodoi{10.1093/mnras/stu1536}

\bibitem[{{Volonteri} \& {Stark}(2011)}]{Volonteri2011}
{Volonteri}, M., \& {Stark}, D.~P. 2011, \mnras, 417, 2085,
  \dodoi{10.1111/j.1365-2966.2011.19391.x}

\bibitem[{{Woo} {et~al.}(2008){Woo}, {Treu}, {Malkan}, \&
  {Blandford}}]{Woo++08}
{Woo}, J.-H., {Treu}, T., {Malkan}, M.~A., \& {Blandford}, R.~D. 2008, \apj,
  681, 925, \dodoi{10.1086/588804}

\bibitem[{{Xue} {et~al.}(2011){Xue}, {Luo}, {Brandt}, {Bauer}, {Lehmer},
  {Broos}, {Schneider}, {Alexander}, {Brusa}, {Comastri}, {Fabian}, {Gilli},
  {Hasinger}, {Hornschemeier}, {Koekemoer}, {Liu}, {Mainieri}, {Paolillo},
  {Rafferty}, {Rosati}, {Shemmer}, {Silverman}, {Smail}, {Tozzi}, \&
  {Vignali}}]{Xue2011}
{Xue}, Y.~Q., {Luo}, B., {Brandt}, W.~N., {et~al.} 2011, \apjs, 195, 10,
  \dodoi{10.1088/0067-0049/195/1/10}

\end{thebibliography}



\end{document}